\documentclass[11pt]{article}
\usepackage{verbatim}
\usepackage[affil-it]{authblk}
\usepackage[latin1]{inputenc}
\usepackage{geometry}
\usepackage{float}
\usepackage{doc}
\usepackage{url}
\usepackage{graphicx}
\usepackage{epstopdf}
\usepackage[T1]{fontenc}
\usepackage{titling}
\usepackage[style=authoryear]{biblatex}
\usepackage{listings}
\bibliography{bibfile}
\title{StegExpose - A Tool for Detecting LSB Steganography}
\author{Benedikt Boehm}
\affil{School of Computing\\ University of Kent, England\\bb269@kent.ac.uk}
\date{}
\begin{document}

\maketitle

\abstract{
Steganalysis tools play an important part in saving time and providing new angles of attack for forensic analysts. StegExpose is a solution designed for use in the real world, and is able to analyse images for LSB steganography in bulk using proven attacks in a time efficient manner. When steganalytic methods are combined intelligently, they are able generate even more accurate results. This is the prime focus of StegExpose.
}

\section{Introduction}
Steganalysis is the practice of detecting the use of steganography. Steganography being the ancient practice of disguising secret communication behind a non suspect channel.\

Proposed here is a steganalysis tool named StegExpose. The tool is built to be universal for detecting steganography in lossless images. StegExpose can be run in the background analysing multiple images without human supervision, returning a detailed steganalytic report once the tool has finished its job.\

The organization of this paper is as follows. Section 2 defines how to interpret specialist terminology in this report. Section 3 reviews adopted technologies and literature. Section 4 discusses the steps taken to create an adequate testing environment for all steganalitic tests. Section 5 covers the attempts to find more accurate and faster steganalysis techniques and presents the test results. Section 6 covers StegExpose's implemented algorithms, features, and usage. Section 7 provides examples of how the tool can be used. Section 8 concludes the project and Section 9 discusses further directions.

\section{Key Terminology}
The following are descriptions of how certain terms are to be understood in the context of this report.
\subsection{LSB steganography, the spatial domain and samples}
LSB stands for 'Least Significant Bit' referring to the bit which makes a byte even or odd. LSB Steganography (also knows as LSB embedding) is a type of digital steganography where secrets are embedded in the least significant bit of a particular sample (or feature) of digital file. The spatial domain refers to a multidimensional space, such as the pixel plane in an image. "Samples" are features within a file that can collectively be used to carry hidden information. In lossless images, the most common samples are individual pixels.
\subsection{Detectors and fusion techniques}
The term detector or signal is used as a shorthand for a steganalytic method. Fusion techniques are a well known concept in signal processing and can be applied to steganalysis. The technique combines multiple detectors into one, with the intention of creating a new detector that is stronger.
\subsection{Stego, carrier, cover and clean files}
Stego files (also knows as carriers) are files that have embedded hidden information as a result of the use of steganography. Covers are files that can potentially be used as carriers (could be any file as long as there exists an embedding method that supports it). Clean files are files that are untouched from steganography.
\subsection{Embedding rate}
The embedding rate refers to the ration between the size of a payload and its cover file. For example if a cover image is 10 MB in size carrying 1 MB of hidden data, the embedding rate of the image would be 10\%.
\subsection{Detector success rate}
Success rate is given a very specific meaning in this report. It refers to the rate at which a particular implementation of a detector is capable of calculating a steganalytic grade for a series of files.

\section{Review of literature and technology}
\subsection{LSB embedding}
In the spatial domain, LSB replacement is the most widely used LSB embedding method. LSB replacement is the process of embedding a secret as-is, so that the secret can be directly read from the LSB's without having to undergo any transformation. More complex LSB embedding methods would obfuscate the payload before embedding it with the intention to make it look statistically like a clean file. Examples include LSB matching \parencite{sharp2001} and Efficient High Payload Data Embedding Scheme or EPES \parencite{omoomi2011}. Keeping a low embedding rate is key in preventing successful steganalysis. This means that it is desirable to embed only into a fraction of all samples (e.g. pixels in images) using a particular distribution method that would decide which sample to use and which to leave out. The importance of keeping embedding rates low is highlighted in \parencite{ker2008}.\
The image embedding tools used in this project are listed below.
\begin{itemize}
  \item LSB-Steganography \parencite{David2012} - LSB replacement with sequential distribution.
  \item OpenStego \parencite{Vaidya2014} - LSB replacement with pseudorandom distribution.
  \item SilentEye \parencite{Chorein2010} - LSB replacement with equidistribution.
  \item OpenPuff \parencite{EmbeddedSW.net2014} - Proprietary method known as "nonlinear adaptive encoding LSB".
\end{itemize}

\subsection{Steganalysis methods}
The following LSB steganalysis methods have been investigated and tested as part of this project. RS analysis \parencite{fridrich2001} detects randomly scattered LSB embedding in grayscale and colour images by inspecting the differences in the number of regular and singular groups for the LSB and 'shifted' LSB plane. Sample pair analysis \parencite{dumitrescu2003} is 'based on a finite state machine whose states are selected multisets of sample pairs called trace multisets' \parencite{dumitrescu2003}. The chi-square attack \parencite{westfeld2000} is a statistical analysis of pairs of values (PoV's) exchanged during LSB embedding. PoV's are groups of binary values within a object's LSB's. Primary sets \parencite{dumitrescu2002} is based on a statistical identity related to certain sets of pixels in an image. The difference histogram analysis \parencite{zhang2003} is a statistical attack on an image's histogram, measuring the correlation between the least significant and all other bit planes.

\subsection{Fusion techniques}
The use of fusion techniques within steganalysis is still largely unexplored. \parencite{kharrazi2006} proved how steganalyis methods can be combined or 'fused' in order to create a stronger detector. Different approaches to fusion are covered such as employing different classification stages and fusion rules. Classification stages include pre and post classification. In pre-classification individual detectors are classified as clean or stego before any further processing is done. In post-classification, various detector outputs (usually percentage values) are combined before classifying an object. Finally, a fusion rule needs to be chosen in order to derive the final indicator. The fusion rule is simply a statistical property that is taken from a set of detectors. \parencite{kharrazi2006} compares and contrasts the mean and maximum rules. More rules are covered by \parencite{kittler1998}, a paper on signal processing, which is also relevant to steganalysis.

\section{Providing a test environment}
In order to achieve quality test results for StegExpose, we generated a pool of 5,200 stego files and 10,000 clean files. All files were sourced from flickr.com, a large image hosting web site. Flickr.com searches were composed of keywords that were likely to return a high diversity of photographic images in terms of colours and textures. Names of countries were most commonly used as keywods. Images in the pool vary in size between 0.04 and 1.02 megapixels, averaging at 0.21. Due to the purpose of flickr.com, most images will be photographic, however non-photographic images will occur on rare occasion.\

Flickr.com hosts only lossy images that are compressed using JPEG. Lossless versions (BMP and PNG) of all images were obtained. After the conversion, images were ready to form part of the pools clean portion. The stego portion had to undergo an embedding operation via SilentEye, OpenStego, OpenPuff or LSB-Steganography where each tool embedded into 1,300 images. All payloads are compressed using the zlib compression library for SilentEye and the .ZIP archive file format for all other stego tools before embedding.\

Stego files created with OpenStego, SilentEye and LSB-Steganography embed the same information into all files, resulting in varying embedding rates, as all carrier files have different sizes. Stego files created with OpenPuff use batch steganography \parencite{ker2007}. Batch steganography is when a single payload is embedded into several files using a uniform embedding rate.\

The table below provides an overview of the resulting embedding rates.
\begin{figure}[H]
    \centering
    \includegraphics[width=0.8\textwidth]{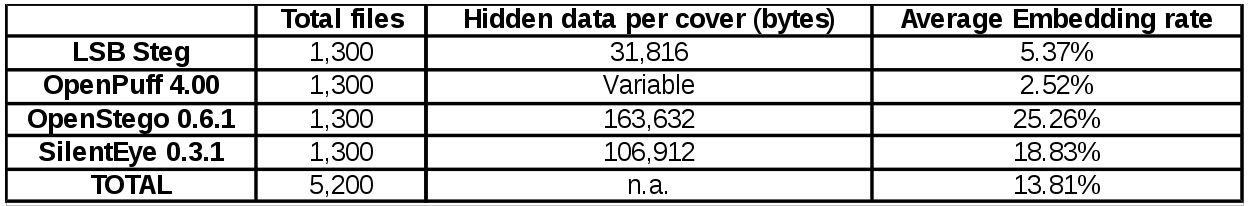}
    \caption{Embedding rates used in the test pool}
    \label{fig:roc}
\end{figure}

\section{Experimentation and Results}

The detectors used in this project were all sourced from other open source steganalysis tools. RS analysis and Sample Pair attack was sourced from Digital Invisible Ink Toolkit \parencite{Hempstalk2006}. Primary Sets and the Chi Square attack were sourced from simple-steganography-suite \parencite{Faure2013}. All detectors are automatic and return a percentage reflecting the likelihood of a file being a carrier.\

The project underwent two rounds of experimentation, namely an accuracy and a speed round. The accuracy round focuses on optimizing the accuracy of a fusion detector, whereas the speed round focuses on finding a detector that provides an acceptable trade-off between accuracy and time. The rounds where necessary in equipping StegExpose with two fusion techniques, standard fusion and fast fusion. The motivation behind this is to make StegExpose relevant for academic as well as practical forensic applications.\

Constants for both rounds include the test pool described in the section 'Providing a test environment'. Additional constants for the speed round include the number of time trials taken by each detector. There are three trials and the average will be used as a speed benchmark. The machine used for running the speed tests will also remain the same. The machine's specifications include a 3.40GHz Intel Core i7-2600 processor with 6 GB of RAM available.

\subsection{Accuracy: finding standard fusion}
The accuracy of all detectors was compared using the area under their ROC (receiver operating characteristic) curve known as the AUC. Where the true positive rate (sensitivity) is plotted over the false positive rate (fall-out). Please note that all AUC values are based on integrating high order polynomial estimates based on 23 ROC coordinates.\

Three different fusion techniques were compared, one which considers only the highest scoring detector, one which considers the arithmetic and one which considers the geometric mean of all detectors.\

\begin{figure}[H]
    \centering
    \includegraphics[width=0.8\textwidth]{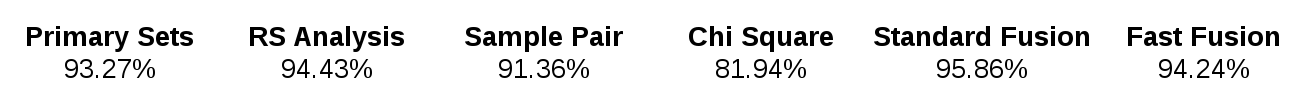}
    \caption{AUC table for fusion detectors and their components}
    \label{fig:auc}
\end{figure}

The arithmetic mean showed the largest AUC and from this point on will be referred to as standard fusion. Standard fusion is more powerful than any of its component detectors, beating runner up RS analysis by 1.43 percentage points in AUC. Figure \ref{fig:roc} shows a ROC curve plotting standard fusion and its component detectors (fast fusion is also plotted and will be discussed in the next section). Figure \ref{fig:auc} gives a table of AUC values, providing a quntitative comparison of all detectors. Fast fusion is also featured in these figures and will be discussed in the next section.
\begin{figure}[H]
    \centering
    \includegraphics[width=0.8\textwidth]{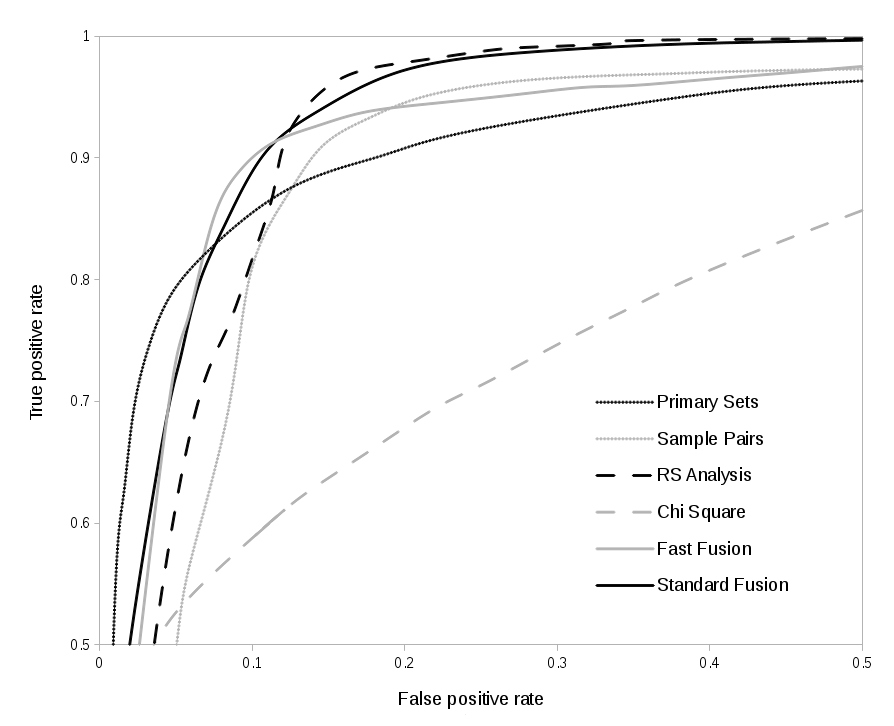}
    \caption{ROC curves for fusion detectors and their components}
    \label{fig:roc}
\end{figure}
Note that if a particular implementation of a detector fails to return a result for a particular file, that file and its detection result is disregarded from the arithmetic mean in the standard fusion algorithm.

\subsection{Speed: finding fast fusion}
The StegExpsose project was interested in finding a second fusion technique that would offer time savings. The technique will be known as fast fusion. Instead of skipping slow detectors completely, StegExpose proposes an algorithms that tries to speed up the classification of clean files, only investing time on suspicious looking ones. This decision was made because in practical applications, clean files are a lot more abundant.\

Any speed results for fast fusion will always be biased towards to the test pool, due to the nature of its algorithm. However, a conservatively high proportion of stego files (a third) in the pool should render results that would rather underestimate the speed of fast fusion.\

\begin{figure}[H]
    \centering
    \includegraphics[width=0.5\textwidth]{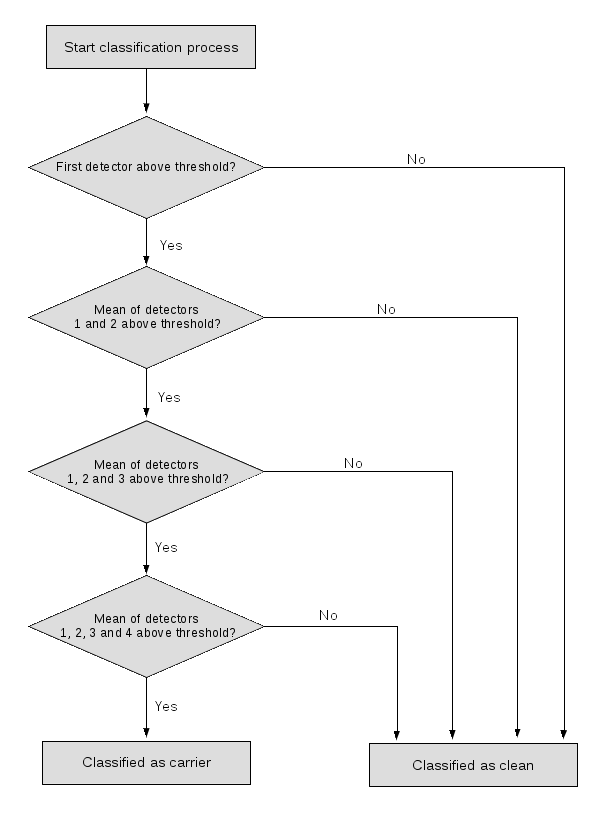}
    \caption{Fast fusion flow chart}
    \label{fig:flowChart}
\end{figure}

\begin{figure}[H]
    \centering
    \includegraphics[width=0.9\textwidth]{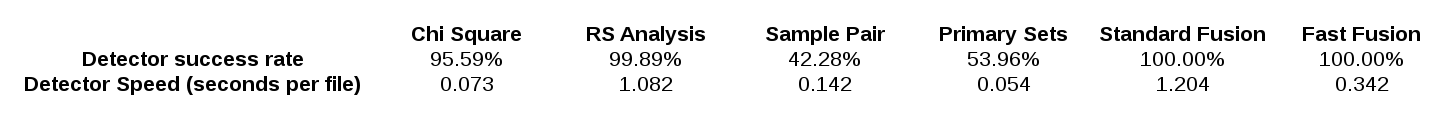}
    \caption{Detector speed table}
    \label{fig:speedTable}
\end{figure}

Fast fusion consists of four stages (one for each component detector). At every stage a new component detector is added and the arithmetic mean of all currently introduced detectors is evaluated. After the evaluation, the result is compared to a specified threshold. If the result is below the threshold, all other stages are skipped and the file is immediately classified as clean. If the result is above threshold, the algorithm passes to the next stage. A file will only be classifies as stego if it passes to the final round and is still above threshold. If a component detector fails to produce a percentage value, the algorithm moves to the next stage giving the failed detector a zero weighting. Figure \ref{fig:flowChart} demonstrates fast fusion's framework in a flow chart.\
\
To complete the described framework, an order of component detectors needs to be established. Initially, the order was solely based on the detector speed. This order proved to be fast but very inaccurate. After testing different orders of component detectors, a particular order proved to be fast as well as accurate. The order takes into consideration the speed as well as the accuracy of the component detectors and goes as follows: 1st Primary Sets, 2nd Sample Pairs, 3rd Chi Square and 4th RS analysis. This order has been chosen for the fast fusion which is 0.19 percentage points (in terms of AUC) less accurate that the strongest component detector (RS analysis), but therefore 3.16 times faster. Figure \ref{fig:auc} and \ref{fig:speedTable} demonstrate fast fusion's accuracy and speed respectively compared to standard fusion and all component detectors.

\section{Implementation and usage of StegExpose}

StegExpose is an open source Java 1.6 program available under https://github.com/b3dk7/StegExpose. There are two main aspects of StegExpose, namely detector fusion and steganalytic reporting.\

\begin{figure}[H]
    \centering
    \includegraphics[width=0.9\textwidth]{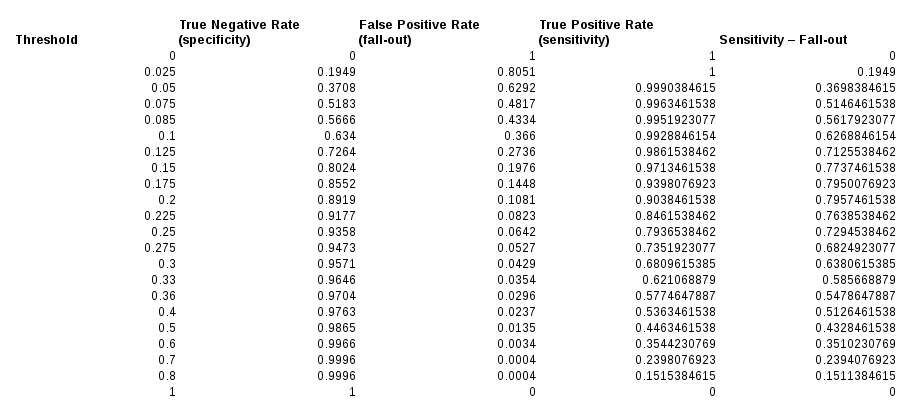}
    \caption{Threshold table for standard fusion}
    \label{fig:thresholdTable}
\end{figure}

\begin{figure}[H]
    \centering
    \includegraphics[width=0.9\textwidth]{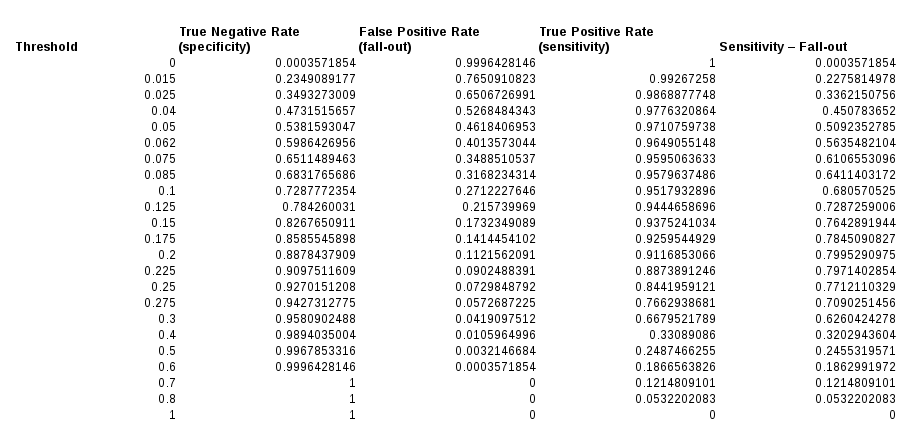}
    \caption{Threshold table for fast fusion}
    \label{fig:thresholdTableFast}
\end{figure}

The detection engine used by StegExpose features standard and fast fusion, which work exactly as described in the previous section. In order to classify images as clean or stego, a threshold must be chosen. A table linking thresholds to ROC values can be found for both fusion detectors under Figure \ref{fig:thresholdTable} and \ref{fig:thresholdTableFast}. From these tables one can gather that the best trade off between fall-out and sensitivity is given at a threshold of 0.2 for both standard and fast fusion. Due to this, StegExpose will use a default threshold of 0.2 unless the user specifies otherwise. Reasons to change the threshold could be to either keep false positives at bay, in which case the threshold would be set slightly higher than the default or to reduce false negatives, in which case the threshold would be set lower. Another benefit of increasing the threshold is that the fusion algorithm will run faster, due to more frequent early classifications taking place. The threshold tables in Figure \ref{fig:thresholdTable} and \ref{fig:thresholdTableFast} can be used for guidance here. Note that for fast fusion, all decision points in Figure \ref{fig:flowChart} use the same threshold i.e. the default or user specified threshold. Both fusion algorithms are implemented as modes i.e. standard and fast mode, collectively known as the fusion modes. A decision needs to be made whether to use standard of fast fusion every time StegExpose is run.\

There are two types of reports that StegExpose is capable of producing, namely the standard and the full report. The standard report prints out to console all files classified as stego and includes an estimate of the size of the embedded data known as quantitative steganalysis. The quantitative steganalysis is derived by mutilpying the fusion detector result by the file size and dividing by three. This method was not included in the 'Experimentation and Results' section, as there was not enough scope to test this thoroughly. However, brief testing showed that the formula is seemingly accurate for covers using embedding rates above 10\%.\

The full report prints out the following information on all files to a csv (comma separated value) file: file name, classification (stego or clean), quantitative steganalysis (payload size in bytes - same technique as for the standard report), Primary Sets result, Chi Square result, Sample Pair result, RS analysis result and fusion result (standard or fast fusion depending on configuration). The steganalityc results for each file are flushed to the report file once fully analysed. This has the effect that any steganalytic progress is not completely lost in case the program crashes.\

In order to run StegExpose, Java 1.6 or later, needs to be installed on the users machine and the StegExpose executable needs to be obtained by creating it from source or directly downloading it from the project repository, where it is saved under 'StegExpose.jar'. Below is an overview of how to run the program and a description of the arguments. Only the first argument is compulsory, however in order to set any of the optional arguments (arguments 2, 3 and 4), all arguments preceding it, must be set.\

\begin{lstlisting}
java -jar StegExpose.jar [directory] [speed] [threshold] [csv file]
\end{lstlisting}

Where

\begin{lstlisting}
[directory]
\end{lstlisting}
Directory containing images to be diagnosed. The directory does not have to exclusively contain images, however only image files will be processed. Beware, that lossy images will be processed as well for which the implemented detectors are not designed.

\begin{lstlisting}
[speed]
\end{lstlisting}
The second argument sets the speed mode. Argument \emph{standard} will run the standard fusion algorithm and \emph{fast} will run the fast fusion algorithm. If the argument is left out, StegExpose will default to standard fusion.

\begin{lstlisting}
[threshold]
\end{lstlisting}
Sets the threshold, taking a floating point value between 0 and 1. If the argument is out of range, not numeric or left out then a default threshold of 0.2 is applied. 

\begin{lstlisting}
[csv file]
\end{lstlisting}
Leaving this argument out will generate a \emph{standard} report outputted to console. Using this argument will generate a \emph{full} report, saving it in the current directory and naming it after the given argument.

\section{Examples of usage}\
Following are some examples in which StegExpose can be used. All examples analyse a directory containing 3 stego files (generated with OpenStego) and 13 clean files available in the project repository under the directory named 'testFolder'.\

\begin{lstlisting}
java -jar StegExpose.jar testFolder
\end{lstlisting}
Basic usage of Stegexpose, providing a directory of images as the only argument. As no other arguments are set, StegExpose defaults to the standard (speed) mode with a threshold of 0.2 and produces the standard report outputted to console.\

\begin{lstlisting}
java -jar StegExpose.jar testFolder standard default steganalysisOfTestFolder
\end{lstlisting}
Same as above but producing a full report named 'steganalysisOfTestFolder' saved under the current directory.\

\begin{lstlisting}
java -jar StegExpose testFolder fast 0.3
\end{lstlisting}
Increasing the threshold and running the program in fast mode.

\section{Conclusion}
StegExpose is a steganalysis tool heavily geared towards bulk analysis of lossless images. Two new fusion detectors, standard and fast fusion were derived from four well known steganalysis methods and successfully implemented in the tool. Standard fusion is more accurate than any of the component detectors it is derived from. Fast fusion is 0.2\% weaker in accuracy than its strongest component but 316\% faster. Note, that these figures are specific to the detector implementations of \parencite{Hempstalk2006} and \parencite{Faure2013} as well as the test pool which has a stego to clean ration of one to three. In a real world setting, the proportion of stego files will be usually a lot lower, causing fast fusion to run even faster.

\section{Further work on StegExpose}
Optimizing quantitative steganalysis in StegExpose is an obvious area for further work, as there has been minimal testing thus far in contrast to it forensic value.\

As written, StegExpose only utilizes one processor core. Featuring multi threading capabilities could significantly increase the speed of running detectors and improve the project as long as it does not introduce any bugs.\

The source code from the project's detector dependencies have remained unchanged. However based on the test pool, the detector success rate (described in the key terminology section) of the implementation of Sample Pair \parencite{Hempstalk2006} and Primary Sets \parencite{Faure2013} analysis is 42\% and 54\% respectively. These figures are very low and are caused by bugs in both dependencies. Fixing these bugs will generate more complete reports and most likely speed up fast mode as well as improve the accuracy of both fusion modes.\

A long term goal for StegExpose would be to introduce image steganalysis in the transform domain (used by the popular JPEG format), as well as other media types such as digital documents, plain text, video and audio. Most importantly, reliable and fast bulk processing needs to be maintained in order to preserve relevance in the practical forensic field.

\section{Acknowledgements}
I would like to thank Julio Hernandez-Castro for supervising this project, proposing the idea behind it and providing invaluable advice. I would also like to thank the authors that proposed the steganalysis methods used in this project as well as Bastien Faure and Kathryn Hempstalk for making their source code freely available.

\printbibliography
\newpage

\end{document}